\documentclass[review]{elsarticle}

\usepackage{stmaryrd}
\usepackage{amsfonts}
\usepackage{graphicx,times,amsmath}
\usepackage{graphics,color}
\usepackage{graphicx}
\usepackage{latexsym}
\usepackage{amsmath}
\usepackage{amsfonts}
\usepackage{amssymb}
\usepackage{url}
\usepackage{subfigure}
\usepackage{times}
\usepackage{color}

\newtheorem{thm}{Theorem}
\newtheorem{lem}{Lemma}
\newtheorem{mydef}{Definition}
\newtheorem{asu}{Assumption}
\newtheorem{col}{Corollary}
\newtheorem{rem}{Remark}
\newtheorem{tion}{Notation}

\allowdisplaybreaks

\allowbreak

\journal{}

\begin{document}

\begin{frontmatter}

\title{Synchronization of linearly coupled reaction-diffusion neural networks with hybrid coupling and time-varying delays via aperiodically intermittent pinning control}

\author[tongji,tjcs]{Xiwei Liu\corref{lxw}}
\ead{xwliu@tongji.edu.cn}
\author[sd]{Zhang Chen}
\ead{chenzhangcz@163.com}
\author[tjmath]{Lingjun Zhou}
\ead{zhoulj@tongji.edu.cn}
\cortext[lxw]{Corresponding author: X. Liu.}

\address[tongji]{Department of Computer Science and
Technology, Tongji University, Shanghai 201804, China}
\address[tjcs]{The Key Laboratory of Embedded System and Service Computing,
Ministry of Education, Shanghai 201804, China}
\address[sd]{School of Mathematics, Shandong University, Jinan 250100, China}
\address[tjmath]{Department of Mathematics, Tongji University, Shanghai 200092, China}

\begin{abstract}
In this paper, the complete synchronization problem of linearly coupled neural networks with reaction-diffusion terms and time-varying delays via aperiodically intermittent pinning control is investigated. The coupling matrix for the network can be asymmetric. Compared with state coupling in the synchronization literature, we design a novel distributed coupling protocol by using the reaction-diffusion coupling-spatial coupling, which can accelerate the synchronization process. This can be regarded as the main difference between this paper and previous works. Using the Lyapunov function and theories in the aperiodically intermittent control, we present some criteria for the complete synchronization with a static coupling strength. In this case, there is no constraint on the bound of time-varying delays, so it can be larger than the length of control span. On the other hand, for the network with an adaptive coupling strength, we propose a simple adaptive rule for the coupling strength and prove its effectiveness rigorously. In this case, the bound of time-varying delay is required to be less than the infimum of the control time span. Finally, numerical simulations are given to verify the theoretical results.
\end{abstract}


\end{frontmatter}


\section{Introduction}
Artificial neural network and its application has been a hot topic in recent decades, including the cellular neural networks, Hopfield neural networks, and Cohen-Grossberg neural networks, etc. Especially, as a great progress in the history of neural networks, Hinton et al. in 2006 proposed the deep learning algorithm \cite{H2006} by using the generalized back-propagation algorithm for training multi-layer neural networks, which has been shown a powerful role in many practical fields, such as natural language processing, image recognition, bioinformatics, recommendation systems, and so on.

Generally, the investigated neural network models are mainly defined by ordinary differential equations (ODEs); however, partial differential equations (PDEs) can describe the real things or events more exactly. For example, diffusion effect cannot be avoided in the neural networks model when electrons are moving in asymmetric electromagnetic field, and in the chemical reactions many complex patterns are generated by the reaction-diffusion effects \cite{Turing,AJB2011}. There have been many results about the dynamical behaviors of neural networks with reaction-diffusion terms, for example, the existence, uniqueness and global exponential stability of the equilibrium point of delayed reaction-diffusion recurrent neural networks were investigated in \cite{LC03}; a better result on stability analysis was given by analyzing the role of the reaction-diffusion terms in \cite{L07}-\cite{WL08}; \cite{Q2007} investigated the exponential stability of impulsive neural networks with time-varying delays and reaction-diffusion terms; \cite{CZ2011} studied the global $\mu$-stability of reaction-diffusion neural networks with unbounded time-varying delays and Dirichlet boundary conditions; and \cite{CFZ2013} discussed the existence and global exponential stability of anti-periodic mild solution for reaction-diffusion Hopfield neural networks systems with time-varying delays.

Neural network can be categorized into the scope of complex networks, while synchronization and its control problem \cite{WL95}-\cite{CLL07} has been an important issue for complex networks, which has attracted many researcher' interests from various disciplines, like mathematics, computer science, physics, biology, etc. Synchronization means that all the nodes have the same dynamical behavior under some local protocols between the communication with each node's neighbours, therefore, synchronization is the collective behavior of the whole network while the communication protocol is distributed. Many synchronization patterns have been investigated, like complete synchronization, cluster synchronization, finite-time (or fixed-time) synchronization, and so on.

In this paper, we will study the complete synchronization and its control problem for linearly coupled neural networks with reaction-diffusion terms. Hitherto, many works have been presented on this problem, see \cite{WC2007}-\cite{WWHR2016}. In \cite{WC2007}, the authors discussed the asymptotic and exponential synchronization for a class of neural networks with time-varying and distributed delays and reaction-diffusion terms; \cite{SYL2009}-\cite{WWG2014} investigated the adaptive synchronization problem for coupled neural networks with reaction-diffusion terms; \cite{liu2010} studied the $\mu$-synchronization problem for coupled neural networks with reaction-diffusion terms and unbounded time-varying delays; \cite{YJ2011}-\cite{SM2012} discussed the synchronization for fuzzy or stochastic neural networks with reaction-diffusion terms and unbounded time delays; \cite{WW2014} proposed a general array model of coupled reaction-diffusion neural networks with hybrid coupling, which was composed of spatial diffusion coupling and state coupling; \cite{WWG2011}-\cite{WWH2015} discussed the synchronization problems of reaction-diffusion neural networks based on the output strict passivity property in which the input and output variables varied with the time and space variables; and \cite{WWHR2016} investigated the pinning control problem for two types of coupled neural networks with reaction-diffusion terms: the state coupling and the spatial diffusion coupling.

As for the control technique, except for the continuous strategy, discontinuous control strategies are more economic and have better application value, including impulsive control, sample-data control, intermittent control, etc. For example, \cite{HJT2010}-\cite{YCY2013} discussed the impulsive control and synchronization for delayed neural networks with reaction-diffusion terms; \cite{RDZ2015} investigated the synchronization problem of reaction-diffusion neural networks with time-varying delays via stochastic sampled-data controller; \cite{HYJT2012}-\cite{MJWLXL2014} studied the synchronization for neural networks with reaction-diffusion terms under periodically intermittent control technique. In the real world, the periodically intermittent control \cite{LiuC11} is rare, and a more general intermittent control technique, called the aperiodically intermittent control, is proposed to realize synchronization of complex networks, see \cite{LC2015,esi2015}. Moreover, the authors in \cite{LC15} and \cite{LLC15} also consider the complete synchronization and the cluster synchronization for complex networks with time-varying delays and constant time delay respectively. Based on the above discussions, in this paper, we will investigate \emph{the complete synchronization problem for linearly coupled neural networks with time-varying delay and reaction-diffusion terms and hybrid coupling via aperiodically intermittent control}.

%

The rest of this paper is organized as follows. In Section \ref{pre}, some necessary definitions, lemmas and notations are given. In Section \ref{main}, the network model with constant aperiodically intermittent control gain is first proposed and the criteria for complete synchronization are obtained. Then, we apply the adaptive approach on the aperiodically intermittent control gain by designing a useful adaptive rule, and its validity is also rigorously proved. In Section \ref{nu}, numerical simulations are presented to illustrate the correctness of theoretical results. Finally, this paper is concluded in Section \ref{conclude}.

\section{Preliminaries}\label{pre}
Some definitions, assumptions, notations and lemmas used throughout the paper will be presented in this section.

At first, we describe the single reaction-diffusion neural network with time-varying delays and
Dirichlet boundary conditions by the following equation:
\begin{align}\label{ys}
\left\{
\begin{array}{ccl}
\frac{\partial{w}_j(t,x)}{\partial{t}}
&=&\sum\limits_{r=1}^m\frac{\partial{}}{\partial{x}_r}\bigg(D_{jr}
\frac{\partial{w}_j(t,x)}{\partial{x}_r}\bigg)
-c_jw_j(t,x)+\sum\limits_{k=1}^na_{jk}g_k(w_k(t,x))\\
&&+\sum\limits_{k=1}^nb_{jk}h_k(w_k(t-\tau(t),x))+\eta_j, \qquad\hfill{} x\in \Omega \\
w_j(t,x)&=&0,\quad\hfill{} (t,x)\in [-\tau,+\infty)\times\partial{\Omega}\\
w_j(t,x)&=&\phi_j(t,x),\quad\hfill{} (t,x)\in
[-\tau,0]\times\Omega
\end{array}\right.
\end{align}
where $j=1,\cdots,n$; $\Omega=\{x=(x_1,x_2,\cdots,x_m)^T||x_r|<l_r, r=1,2,\cdots,m\}$ is a compact set with smooth boundary $\partial{\Omega}$ and mes$\Omega>0$ in space $R^m$; $w_j(t,x)\in R$ is the state of $j$th neuron at time $t$ and in space
$x$; $D_{jr}\ge 0$ means the transmission diffusion coefficient along the $j$th neuron; $c_j>0$ represents the rate with which the $j$th neuron will reset its potential to the resting state; $a_{jk}$ and $b_{jk}$ are the connection weights with and without delays respectively; $g_k(t,x)$ and $h_k(t,x)$ denote the activation functions of the $j$th neuron in space $x$; $\eta_j$ denotes the external bias on the $j$th neuron; $\tau(t)$ is the bounded time-varying delay with $\tau(t)\le \tau$, where $\tau>0$; the initial values $\phi_j(t, x)$ are bounded and continuous functions.

Denote
\begin{align*}
&C=\mathrm{diag}(c_1,\cdots,c_n)\in R^{n\times n};\\
&A=(a_{jk})\in R^{n\times n}; B=(b_{jk})\in R^{n\times n};\\
&\eta=(\eta_1,\cdots,\eta_n)^T\in R^n;\\
&w(t,x)=(w_1(t,x),\cdots,w_n(t,x))^T\in R^n;\\
&\frac{\partial{w}(t,x)}{\partial{t}}=\bigg(\frac{\partial{w}_1(t,x)}{\partial{t}},\cdots,
\frac{\partial{w}_n(t,x)}{\partial{t}}\bigg)^T\in R^n;\\
&g(w(t,x))=(g_1(w_1(t,x)),\cdots,g_n(w_n(t,x)))^T\in R^n;\\
&h(w(t,x))=(h_1(w_1(t,x)),\cdots,h_n(w_n(t,x)))^T\in
R^n;\\
&f^{rd}(w(t,x))\\
&=\Bigg(\sum\limits_{r=1}^m\frac{\partial{}}{\partial{x}_r}
\bigg(D_{1r}\frac{\partial{w}_1(t,x)}{\partial{x}_r}\bigg),\cdots,
\sum\limits_{r=1}^m\frac{\partial{}}{\partial{x}_r}
\bigg(D_{nr}\frac{\partial{w}_n(t,x)}{\partial{x}_r}\bigg)
\Bigg)^T\in R^n;\\
&f(w(t,x);w(t-\tau(t),x))\\
&=f^{rd}(w(t,x))-Cw(t,x)+Ag(w(t,x))+Bh(w(t-\tau(t),x))+\eta\in R^n.
\end{align*}
With these notations, (\ref{ys}) can be rewritten in the compact
form as
\begin{align}\label{ys2}
\frac{\partial{w}(t,x)}{\partial{t}}=f(w(t,x);w(t-\tau(t),x)),
\end{align}
with the norm $\|\cdot\|$ defined by
\begin{align}
\|w(t,x)\|=\bigg[\int_{\Omega}w(t,x)^Tw(t,x)dx\bigg]^{\frac{1}{2}}.
\end{align}

As for the activation functions $g(\cdot)$ and $h(\cdot)$,
\begin{asu}\label{add}
There exist positive constants $g^{\star}$ and $h^{\star}$, such that for any vectors $w(t,x)$ and $\pi(t,x)\in R^m$, the following conditions hold:
\begin{align*}
\|g(w(t,x))-g(\pi(t,x))\|\le
g^{\star}\|(w(t,x)-\pi(t,x))\|,\\
\|h(w(t,x))-h(\pi(t,x))\|\le h^{\star}\|(w(t,x)-\pi(t,x))\|.
\end{align*}
\end{asu}

As for the effect of reaction-diffusion, the following lemma plays an important role.
\begin{lem}\label{use} (See \cite{Lu2008})
Let $\Omega$ be a cube $|x_r|<l_r, r=1,2,\cdots,m$, and let $q(x)$ be a real-valued function belonging to $C^1(\Omega)$ which vanishes on the boundary $\partial{\Omega}$, i.e., $q(x)|_{\partial{\Omega}}=0$. Then
\begin{align*}
\int_{\Omega}q^2(x)dx\le l_r^2\int_{\Omega}\bigg{|}\frac{\partial{q(x)}}{\partial{x_r}}\bigg{|}^2dx.
\end{align*}
\end{lem}

For a complex network with $N$ linearly coupled neural networks (\ref{ys2}), we suppose its coupling configuration (or the corresponding coupling matrix $\Xi=(\xi_{jk})\in R^{N\times N}$) satisfies that:
\begin{enumerate}
  \item $\xi_{jk}\geq 0, j\neq k, \xi_{jj}=-\sum_{k=1, k\neq j}^N \xi_{jk}, j=1,\cdots,N$;
  \item $\Xi$ is irreducible.
\end{enumerate}

\begin{lem}\label{lyasta}
(See \cite{CLL07}) For the above $N\times N$ coupling matrix $\Xi$ and a constant $\sigma>0$, the new matrix
\begin{align}\label{newmatrix}
\tilde{\Xi}=\Xi-\mathrm{diag}(\sigma,0,\cdots,0)
\end{align}
is Lyapunov stable. Especially, suppose the left eigenvector corresponding to the eigenvalue `$0$' of matrix $\Xi$ is $p=(p^1,\cdots,p^N)^T$, then $P=\mathrm{diag}(p^1,\cdots,p^N)>0$ satisfies the following condition
\begin{align*}
P\tilde{\Xi}+\tilde{\Xi}^TP<0.
\end{align*}
Without loss of generality, in the following, we assume that $\max_{j=1,\cdots,N}p^j=1$.
\end{lem}

In this paper, we adopt the aperiodically intermittent control strategy firstly proposed in \cite{LC2015}-\cite{LLC15}, whose description can be found in Figure 1. For the $i$-th time span $[t_i, t_{i+1})$, it is composed of control time span $[t_i, s_i]$ and rest time span $(s_i, t_{i+1})$, where $t_0=0$. Especially, when $s_i-t_i\equiv\theta$ and $t_{i+1}-t_i\equiv\omega$, where constants $\theta, \omega\ge 0$, then the aperiodically intermittent control becomes the periodically intermittent control, see \cite{HYJT2012}-\cite{LiuC11} and reference therein.

\begin{figure}
\begin{center}
\includegraphics[width=\textwidth,height=0.23\textheight]{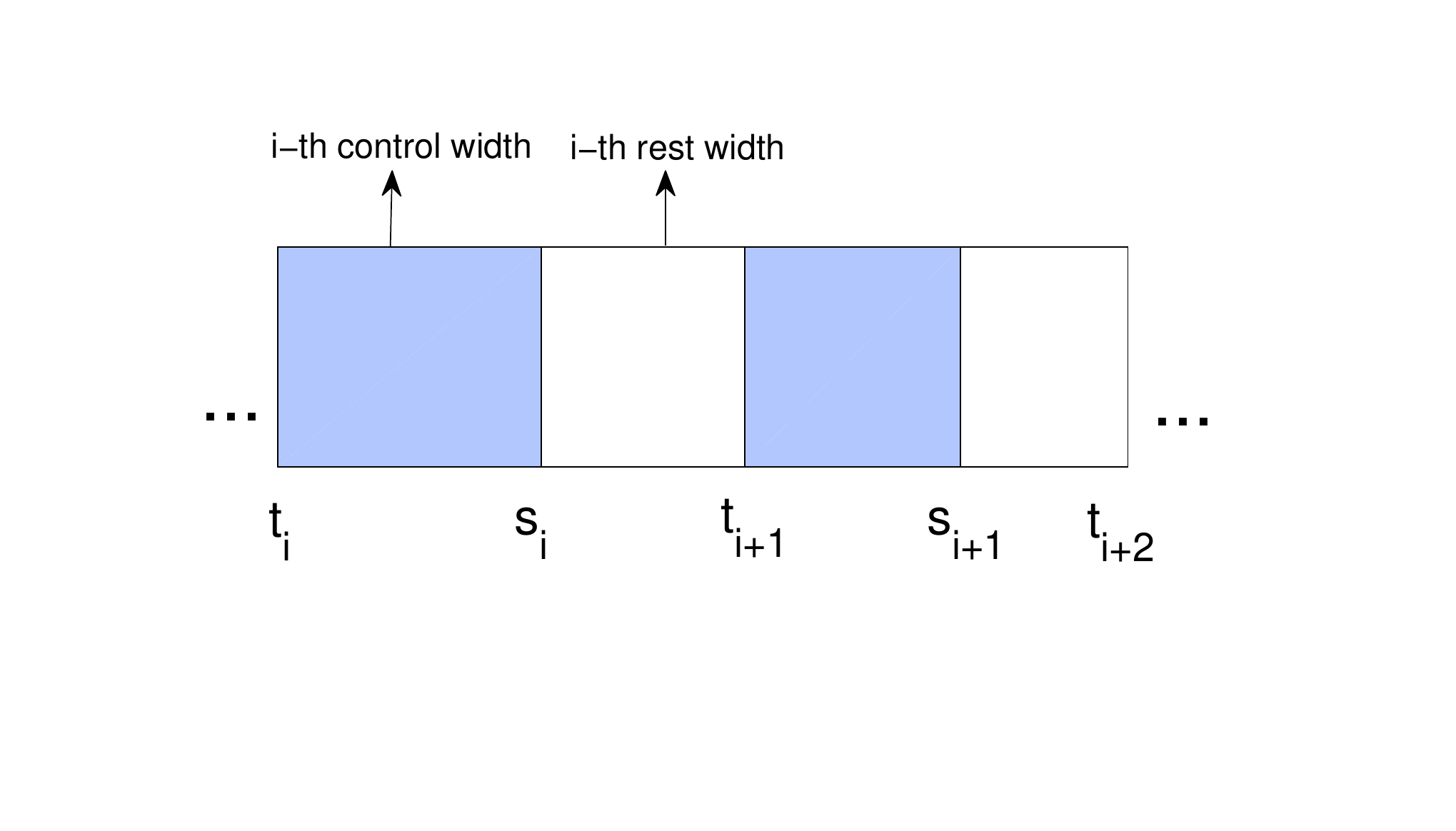}
\end{center}
\label{sk}
\caption{Sketch map of the aperiodically intermittent control strategy}
\end{figure}

For the aperiodically intermittent control strategy in Figure 1, some important definitions are given.
\begin{mydef}\label{rho}
(\cite{LC2015}-\cite{LLC15}) The proportion of the rest width $t_{i+1}-s_i$ in the time span $t_{i+1}-t_i$ is define by
\begin{align}\label{index}
\rho(t)=\limsup\limits_{l\to \infty}\frac{t-s_i}{t-t_i}, ~~\mathrm{where}~~ t\in [s_i, t_{i+1}].
\end{align}
Obviously, $\rho(t)$ is an increasing function, and the maximum value is achieved when $t=t_{i+1}$. Without loss of generality, we always suppose that $\rho(t_{i+1})\in (0,1)$.
\end{mydef}

The following assumption is usually used in the aperiodically intermittent control strategy.
\begin{asu}\label{asutime}
(\cite{LC2015}-\cite{LLC15}) There exist two positive scalars $0<\theta<\omega<+\infty$, such that
\begin{align*}
\inf\limits_{i=0,1,2,\cdots}(s_i-t_i)=\theta, ~~\mathrm{and}~~\sup\limits_{i=0,1,2,\cdots}(t_{i+1}-t_i)=\omega.
\end{align*}
\end{asu}
This assumption means in each time span $[t_i, t_{i+1})$, the control width is no less than $\theta$, and the rest width should be no more than $\omega-\theta$.

\begin{rem}
In the literature of synchronization under aperiodically intermittent control with time delays, the comparison among the rest width, the control width and the largest time delay $\tau$ is an important issue, see \cite{LC15}, which will also be carefully discussed in Section \ref{main}.
\end{rem}

\begin{rem}
Under Assumption \ref{asutime}, one can easily get that the maximum value of index function $\rho(t)$ defined in (\ref{index}) is:
\begin{align}\label{max}
\rho^{\star}=\max_i\rho(t_{i+1})=1-\frac{\theta}{\omega}.
\end{align}
\end{rem}


The following lemma is proposed to deal with the stability problem with aperiodically intermittent control and time-varying delays.

\begin{lem}\label{synlem}
(\cite{LC15}) Assume the function $V(t)$ is continuous and nonnegative when $t\in [-{\tau},+\infty)$, and satisfies the following condition:
\begin{align}\label{del}
\left\{
\begin{array}{ll}
\dot{V}(t)\leq -\beta_1V(t)+\beta_2V(t-\tau(t)),&t_i\leq t\leq s_i,\\
\dot{V}(t)\leq \beta_3V(t)+\beta_2V(t-\tau(t)),&s_i\leq t\leq t_{i+1},
\end{array}
\right.
\end{align}
where $\beta_1, \beta_2, \beta_3$ are constants, and $i=0,1,2,\cdots$.  Suppose that for the aperiodically intermittent pinning control, there exists a constant $\rho^{\star}\in (0,1)$ defined in (\ref{max}). If
\begin{align*}
\beta_1>\beta_2\geq 0,~~\beta=\beta_1+\beta_3>0, ~~\varepsilon=\lambda-\beta\rho^{\star}>0,
\end{align*}
then
\[
V(t)\leq \sup_{-{\tau}\leq \kappa \leq 0}V(\kappa)\mathrm{exp}\{-\varepsilon t\}, t \geq 0,
\]
where $\lambda>0$ is the unique positive solution of the equation
\begin{align*}
\lambda-\beta_1+\beta_2 \mathrm{exp}\{\lambda{\tau}\}=0.
\end{align*}
\end{lem}

\begin{rem}
There is no constraint on the constant ${\tau}$, which can be large enough to surpass the rest width $t_{i+1}-s_i$. This Lemma makes our obtained results be applied in a wider scope. If the time delay is constant, then \cite{LLC15} presents a more conservative result, interested readers are encouraged to investigate the synchronization problem under constant time delay.
\end{rem}

\begin{tion}
For the rest of this paper, we use
$I$ to denote the identity matrix with appropriate dimension, the symmetric part of matrix $A$ is defined as $A^s=(A+A^T)/2$. A symmetric real matrix $A$ is positive definite
(semi-definite) if $x^T Ax>0 (\geq 0)$ for all nonzero vectors $x$, denoted as $A>0 (A\geq 0)$. $\lambda_{\min}(A)$ and $\lambda_{\max}(A)$ mean the minimum and maximum eigenvalues of symmetric matrix $A$ respectively. `$A\otimes B=(a_{ij}B)$' is the Kronecker product of two matrices $A$ and $B$.
\end{tion}


\section{Main results}\label{main}
In this section, we first give the network model of linearly coupled neural networks with reaction-diffusion terms and time-varying delays via aperiodically intermittent pinning control. Especially, the coupling is hybrid coupling composed of spatial diffusion coupling and state coupling, which was proposed and investigated in \cite{WW2014,WWHR2016}.

\subsection{Network model}
Without loss of generality, we assume that only
the first node (neural network) is pinned with a constant control gain, so the network model can be described as
\begin{align}\label{m-1}
\left\{
\begin{array}{cl}
\frac{\partial{w^1(t,x)}}{\partial{t}}=&f(w^1(t,x);w^1(t-\tau(t),x))\\
&+\sum\limits_{k=1}^N\xi_{1k}\Gamma^1
w^k(t,x)-\sum\limits_{k=1}^N\xi_{1k}\Gamma^2
\Delta(w^k(t,x))+u^1(t,x),\\
&~\mathrm{if}~~t\in [t_i,s_i], i=0,1,2,\cdots\\
\frac{\partial{w^j(t,x)}}{\partial{t}}=&f(w^j(t,x);w^j(t-\tau(t),x))\\
&+\sum\limits_{k=1}^N\xi_{jk}\Gamma^1
w^k(t,x)-\sum\limits_{k=1}^N\xi_{jk}\Gamma^2
\Delta(w^k(t,x)),\\
&j=1,2,\cdots,N
\end{array}
\right.
\end{align}
where $w^j(t,x)\in R^n$ is the state vector of the $j$th neural network; $f(\cdot;\cdot)$ denotes the original dynamical behavior of uncoupled neural networks defined in (\ref{ys2}); the term $\sum\limits_{k=1}^N\xi_{jk}\Gamma^1
w^k(t,x)$ is the common linearly coupled term between nodes using the states, while the term $-\sum\limits_{k=1}^N\xi_{jk}\Gamma^2
\Delta(w^k(t,x))$ is the spatial diffusion coupling term, where $\Delta=\sum_{r=1}^m{\partial^2}/{\partial{x_r^2}}$ is the Laplace diffusion operator on $\Omega$; the inner coupling matrices $\Gamma^1=\mathrm{diag}(\gamma_1^1,\cdots,\gamma_n^1)$ and $\Gamma^2=\mathrm{diag}(\gamma_1^2,\cdots,\gamma_n^2)$ are diagonal; $t_i$ and $s_i$ are defined by the aperiodically intermittent control in Figure 1.

The initial values for the delayed network (\ref{m-1}) are given by
\begin{align*}
w^j(t,x)=(\phi^j_1(t,x),\cdots,\phi^j_n(t,x))^T\in C([-{\tau},0]\times R^m,R^n), ~~j=1,2,\cdots,N,
\end{align*}
where $C([-{\tau},0]\times R^m,R^n)$ denotes the Banach space of all continuous functions mapping $[-{\tau}, 0]\times R^m$ into $R^n$.

Suppose the target trajectory is $\pi(t,x)$, whose equation is governed by
\begin{align}\label{origin}
\frac{\partial{\pi}(t,x)}{\partial{t}}=f(\pi(t,x);\pi(t-\tau(t),x)),
\end{align}
where $\Omega$ is the cube with $|x_r|<l_r, r=1,2,\cdots,m$.

Then, for the pinning control $u^1(t,x)$, we can design its form as:
\begin{align}\label{controlnew}
u^1(t,x)=\sigma\Gamma^1 (\pi(t,x)-w^1(t,x))-\sigma\Gamma^2 \Delta(\pi(t,x)-w^1(t,x)),
\end{align}
where $\sigma$ is the pinning control gain.

\begin{rem}
Of course, only the state diffusion (the first term in (\ref{controlnew})) or the spatial diffusion (the second term in (\ref{controlnew})) can also be used as the pinning control technique, see \cite{WWHR2016}, but in order to consider them together, we adopt the form of (\ref{controlnew}). As for the negative sign for the spatial diffusion coupling term, the reason is that this form can be helpful for the synchronization through our following proof process.
\end{rem}

\begin{rem}
In the network model (\ref{m-1}), we assume that the outer coupling matrices are the same for state coupling term and spatial diffusion coupling term. There are two reasons, one is that the coupling configuration denotes the information interchange between nodes, therefore, a node has the same communication with its neighbours for both state coupling and spatial diffusion coupling; the other one is that this assumption can make the following calculation clear to understand. The same reason is also for the control gain $\sigma$.
\end{rem}

\subsection{Complete synchronization with a static coupling strength}\label{static}
For any node $j=1,2,\cdots,N$, denote $e^j(t,x)=(e^j_1(t,x),\cdots,e^j_n(t,x))^T=w^{j}(t,x)-\pi(t,x)\in R^n$ as the synchronization error, then the aim of this paper is to investigate under what conditions the complete synchronization can be realized, i.e.,
\begin{align}
\lim_{t\rightarrow +\infty}\|e(t,\cdot)\|=0,
\end{align}
where
$e(t,x)=(e^1(t,x)^T,\cdots,e^N(t,x)^T)^T$, and
\begin{align}\label{norm}
\|e(t,\cdot)\|=\sqrt{\sum_{j=1}^N\|e^j(t,\cdot)\|^2}=\sqrt{\sum_{j=1}^N\int_{\Omega}e^j(t,x)^Te^j(t,x)dx}.
\end{align}

Denote
\begin{align*}
&f^{rd}(e^j(t,x))=\Bigg(\sum\limits_{r=1}^m\frac{\partial{}}{\partial{x}_r}
\bigg(D_{1r}\frac{\partial{e}_1^j(t,x)}{\partial{x}_r}\bigg),\cdots,
\sum\limits_{r=1}^m\frac{\partial{}}{\partial{x}_r}
\bigg(D_{nr}\frac{\partial{e}_n^j(t,x)}{\partial{x}_r}\bigg)
\Bigg)^T;\\
&\tilde{g}(e^j(t,x))=g(w^j(t,x))-g(\pi(t,x));\\
&\tilde{h}(e^j(t,x))=h(w^j(t,x))-h(\pi(t,x));\\
&\tilde{f}(e^j(t,x);e^j(t-\tau(t),x))\nonumber\\
&=f^{rd}(e^j(t,x))-Ce^j(t,x)+A\tilde{g}(e^j(t,x))+B\tilde{h}(e^j(t-\tau(t),x)).
\end{align*}
Then the dynamics of error $e^j(t,x)$ can be described as:
\begin{align}\label{m-2}
\left\{
\begin{array}{cl}
\frac{\partial{e^1(t,x)}}{\partial{t}}=&\tilde{f}(e^1(t,x);e^1(t-\tau(t),x))\\
&+\sum\limits_{k=1}^N\tilde{\xi}_{1k}\Gamma^1
e^k(t,x)-\sum\limits_{k=1}^N\tilde{\xi}_{1k}\Gamma^2
\Delta(e^k(t,x)),\\
&~\mathrm{if}~~t\in [t_i,s_i], i=0,1,2,\cdots\\
\frac{\partial{e^j(t,x)}}{\partial{t}}=&\tilde{f}(e^j(t,x);e^j(t-\tau(t),x))\\
&+\sum\limits_{k=1}^N\xi_{jk}\Gamma^1
e^k(t,x)-\sum\limits_{k=1}^N\xi_{jk}\Gamma^2
\Delta(e^k(t,x)),\\
&j=1,2,\cdots,N
\end{array}
\right.
\end{align}
where
\[
\tilde{\xi}_{jk}=\left\{\begin{array}{ll}\xi_{11}-\sigma, &j=k=1;\\
\xi_{jk}, &\mathrm{otherwise}.\end{array}\right.
\]

\begin{thm}\label{thm1}
For the network (\ref{m-1}), suppose Assumption \ref{add} holds, $\Gamma^1$ and $\Gamma^2$ are all positive definite matrices. Denote
\begin{align}
&d=\min_{k=1,\cdots,n}\sum_{r=1}^m\frac{D_{kr}}{l_r^2};\label{wa1}\\
&\alpha_1=2\lambda_{\max}(-C+\varepsilon_1^{-1}AA^T+\varepsilon_1{g^{\star}}^2+\varepsilon_2^{-1}BB^T);\label{w2}\\
&\alpha_2=2\varepsilon_2{h^{\star}}^2;\label{w3}\\
&\alpha_3=2\min_{k=1,\cdots,n}\gamma_k^1\lambda_{\max}(\{P\tilde{\Xi}\}^s);\label{w4}\\
&\alpha_4=2\min_{k=1,\cdots,n}\gamma_k^2\lambda_{\max}(\{P\tilde{\Xi}\}^s)\sum_{r=1}^m\frac{1}{l_r^2};\label{w5}\\
&\beta_1=2d-\alpha_1-\alpha_3-\alpha_4;\label{w6}\\
&\beta_3=\alpha_1-2d;\label{w7}\\
&\beta=\beta_1+\beta_3=-\alpha_3-\alpha_4;\label{w8}
\end{align}
where $\tilde{\Xi}$ and $P$ are defined in Lemma \ref{lyasta}, $\varepsilon_1$ and $\epsilon_2$ are positive constants. Obviously, $\alpha_3<0$, $\alpha_4<0$, and $\beta>0$. Moreover, for the aperiodically intermittent pinning control, suppose Assumption \ref{asutime} holds and there exists a constant $\rho^{\star}\in (0,1)$ defined in (\ref{max}). Assume $\beta_1>\alpha_2\geq 0$, then their exists a unique positive solution $\lambda$ of the equation
\begin{align*}
\lambda-\beta_1+\alpha_2 \mathrm{exp}\{\lambda{\tau}\}=0.
\end{align*}
If $\delta=\lambda-\beta\rho^{\star}>0$, then the complete synchronization can be realized exponentially, i.e., $\|e(t,\cdot)\|=O(e^{-\delta t/2})$.
\end{thm}
The proof will be given in Appendix A.

\begin{rem}\label{lmi}
As for the calculation of parameters $\epsilon_1$ and $\epsilon_2$, one can use the Linear Matrix Inequality (LMI) technique to solve, since LMI has been widely used in the literature of neural networks, here we omit the details. Moreover, as for the relations between $\alpha_3$ and $\rho^{\star}$, \cite{LL16} gives a careful discussion.
\end{rem}

\begin{rem}\label{red}
From the process of proof, one can find that the diffusive term $f^{rd}(w^j(t,x))$ is beneficial for the complete synchronization, whose role can be expressed by the parameter $d$. When $\beta$ and $\rho^{\star}$ are fixed, along with the increase of $d$, $\beta_1$ and $\lambda$ increase, the index $\delta$ will also increase, which means that the synchronization can be achieved with a faster speed. Especially, when the diffusive term $f^{rd}(w^j(t,x))$ vanishes, the synchronization speed is the smallest.
\end{rem}

\begin{rem}
The spatial diffusion coupling term  $-\sum\limits_{k=1}^N\xi_{jk}\Gamma^2
\Delta(w^k(t,x))$ is helpful for the synchronization from the proof, this is the reason why we design the spatial diffusion coupling term as this form. On the other hand, the normal coupling form $\sum\limits_{k=1}^N\xi_{jk}\Gamma^2
\Delta(w^k(t,x))$ may play a negative role for synchronization. The concept of spatial diffusion coupling was first proposed by works \cite{WW2014,WWHR2016}. In these papers, the coupling term is chosen as the normal coupling form as $\sum\limits_{k=1}^N\xi_{jk}\Gamma^2
\Delta(w^k(t,x))$. Therefore, in their handling with the spatial diffusion coupling term, they used the original reaction-diffusion term $f^{rd}(w^j(t,x))$ discussed in Remark \ref{red} to suppress the negative effect of the spatial diffusion coupling term (see (38) in \cite{WWHR2016}), while in this paper the spatial diffusion coupling term can help to facilitate the synchronization, this can be regarded as the main difference between this paper with previous works.
\end{rem}

\begin{rem}
In some cases, the parameters in the network model are unknown, or the real required coupling strength is smaller than the calculated value by Theorem \ref{thm1}. Therefore, a method is to apply the adaptive technique, which is also investigated in many papers, and according to the analysis in Remark 3 of \cite{LC15}, the adaptive rule for this case (large delay) cannot be rigorously proved. In subsection \ref{adaptive}, we will give more discussions about small delay case.
\end{rem}

In the following, we present some simple corollaries, which can be directly obtained from Theorem \ref{thm1}.
\begin{col}
For the coupled neural networks only with the state coupling,
\begin{align*}
\left\{
\begin{array}{cl}
\frac{\partial{w^1(t,x)}}{\partial{t}}=&f(w^1(t,x);w^1(t-\tau(t),x))+\sum\limits_{k=1}^N\xi_{1k}\Gamma^1
w^k(t,x)+u^1(t,x)\\
&+\sigma\Gamma^1 (\pi(t,x)-w^1(t,x)),\\
&~\mathrm{if}~~t\in [t_i,s_i], i=0,1,2,\cdots\\
\frac{\partial{w^j(t,x)}}{\partial{t}}=&f(w^j(t,x);w^j(t-\tau(t),x))+\sum\limits_{k=1}^N\xi_{jk}\Gamma^1
w^k(t,x), j=1,2,\cdots,N
\end{array}
\right.
\end{align*}
where $\pi(t,x)$ is defined in (\ref{origin}), suppose Assumption \ref{add} holds, and the notations $d$, $\alpha_1$, $\alpha_2$, $\alpha_3$ and $\beta_3$ are defined in (\ref{wa1})-(\ref{w4}) and (\ref{w7}). Denote
\begin{align*}
\beta_1=2d-\alpha_1-\alpha_3, \beta=\beta_1+\beta_3=-\alpha_3>0,
\end{align*}
where $\tilde{\Xi}$ and $P$ are defined in Lemma \ref{lyasta}, $\varepsilon_1$ and $\epsilon_2$ are positive constants. For the aperiodically intermittent pinning control, suppose Assumption \ref{asutime} holds and there exists a constant $\rho^{\star}\in (0,1)$ defined in (\ref{max}). Assume $\beta_1>\alpha_2\geq 0$, then their exists a unique positive solution $\lambda$ of the equation
$\lambda-\beta_1+\alpha_2 \mathrm{exp}\{\lambda{\tau}\}=0$.
If $\delta=\lambda-\beta\rho^{\star}>0$, then the complete synchronization can be realized exponentially, i.e., $\|e(t,\cdot)\|=O(e^{-\delta t/2})$.
\end{col}

\begin{col}
For the coupled neural networks only with the spatial diffusion coupling,
\begin{align}\label{c-2}
\left\{
\begin{array}{cl}
\frac{\partial{w^1(t,x)}}{\partial{t}}=&f(w^1(t,x);w^1(t-\tau(t),x))-\sum\limits_{k=1}^N\xi_{1k}\Gamma^2
\Delta(w^k(t,x))\\
&-\sigma\Gamma^2 \Delta(\pi(t,x)-w^1(t,x)),\\
&~\mathrm{if}~~t\in [t_i,s_i], i=0,1,2,\cdots\\
\frac{\partial{w^j(t,x)}}{\partial{t}}=&f(w^j(t,x);w^j(t-\tau(t),x))-\sum\limits_{k=1}^N\xi_{jk}\Gamma^2
\Delta(w^k(t,x)),\\
&j=1,2,\cdots,N
\end{array}
\right.
\end{align}
where $\pi(t,x)$ is defined in (\ref{origin}), suppose Assumption \ref{add} holds, and the notations $d$, $\alpha_1$, $\alpha_2$, $\alpha_4$ and $\beta_3$ are defined in (\ref{wa1})-(\ref{w3}), (\ref{w5}) and (\ref{w7}). Denote
\begin{align*}
\beta_1=2d-\alpha_1-\alpha_4, \beta=\beta_1+\beta_3=-\alpha_4>0,
\end{align*}
where $\tilde{\Xi}$ and $P$ are defined in Lemma \ref{lyasta}, $\varepsilon_1$ and $\epsilon_2$ are positive constants. For the aperiodically intermittent pinning control, suppose Assumption \ref{asutime} holds and there exists a constant $\rho^{\star}\in (0,1)$ defined in (\ref{max}). Assume $\beta_1>\alpha_2\geq 0$, then their exists a unique positive solution $\lambda$ of the equation
$\lambda-\beta_1+\alpha_2 \mathrm{exp}\{\lambda{\tau}\}=0$.
If $\delta=\lambda-\beta\rho^{\star}>0$, then the complete synchronization can be realized exponentially, i.e., $\|e(t,\cdot)\|=O(e^{-\delta t/2})$.
\end{col}

\subsection{Complete synchronization with an adaptive coupling strength}\label{adaptive}
In \cite{LC15}, the authors proposed an adaptive rule for synchronization under aperiodically intermittent pinning control when the time-varying delay is very small. In this paper, we will adopt this approach to solve our problem. In this case, we assume
\begin{align}\label{smalltau}
\tau(t)\leq {\tau}<\inf\limits_i\{s_i-t_i\}.
\end{align}

\begin{lem}\label{ht}
(\cite{LC15}) Suppose
\begin{align*}
\dot{V}(t)=M(V(t), V(t-\tau(t)))-\Psi(t)V(t),
\end{align*}
where $V(t)\in R, M(\cdot, \cdot): R\times R\rightarrow R$ is continuous and satisfies
\begin{align*}
V(t)\cdot M(V(t), V(t-\tau(t)))\leq L_1V^2(t)+L_2V^2(t-\tau(t)),
\end{align*}
where $L_1>0, L_2>0$ and $M(0,0)=0$. Function $\Psi(t):{0\bigcup R_+}\rightarrow R_+$ is the adaptive aperiodically intermittent feedback control gain defined as follows:
\begin{align}\label{good}
\Psi(t)=\left\{
\begin{array}{lll}
0; & t=0\\
0; & t\in (s_i,t_{i+1}),&i=0,1,2,\cdots\\
\Psi(s_i); &t=t_{i+1},&i=0,1,2,\cdots
\end{array}
\right.
\end{align}
and
\begin{align}
\dot{\Psi}(t)=\psi\max\limits_{-{\tau}\leq \varsigma\leq 0}V^2(t+\varsigma), \psi>0, t\in [t_i, s_i], i=0,1,2,\cdots
\end{align}
If (\ref{smalltau}) holds, then $\lim_{t\rightarrow \infty}V(t)=0$.
\end{lem}

Now, the network (\ref{m-1}) with an adaptive coupling strength via aperiodically intermittent control can be modelled as follows:
\begin{align}\label{a-1}
\left\{
\begin{array}{cl}
\frac{\partial{w^1(t,x)}}{\partial{t}}=&f(w^1(t,x);w^1(t-\tau(t),x))\\
&+\Psi(t)\sum\limits_{k=1}^N\xi_{1k}\Gamma^1
w^k(t,x)-\Psi(t)\sum\limits_{k=1}^N\xi_{1k}\Gamma^2
\Delta(w^k(t,x))\\
&+\Psi(t)\sigma\Gamma^1 (\pi(t,x)-w^1(t,x))-\Psi(t)\sigma\Gamma^2 \Delta(\pi(t,x)-w^1(t,x)),\\
\frac{\partial{w^j(t,x)}}{\partial{t}}=&f(w^j(t,x);w^j(t-\tau(t),x))\\
&+\Psi(t)\sum\limits_{k=1}^N\xi_{jk}\Gamma^1
w^k(t,x)-\Psi(t)\sum\limits_{k=1}^N\xi_{jk}\Gamma^2
\Delta(w^k(t,x)), \\
&~~~~~~~~~~~~~~~~~~~~~~~~~~~~~~~~~~~~~~~~~~~~~~~~~~~~~~~~~~~~~~~~~~~~~~~~~~~~~j=2,3,\cdots,N,\\
&~\mathrm{if}~~t\in [t_i,s_i], i=0,1,2,\cdots\\
\frac{\partial{w^j(t,x)}}{\partial{t}}=&f(w^j(t,x);w^j(t-\tau(t),x))\\
&+\sum\limits_{k=1}^N\xi_{jk}\Gamma^1
w^k(t,x)-\sum\limits_{k=1}^N\xi_{jk}\Gamma^2
\Delta(w^k(t,x)), ~j=1,2,\cdots,N,\\
&~\mathrm{if}~~t\in (s_i,t_{i+1}), i=0,1,2,\cdots
\end{array}
\right.
\end{align}
where $\pi(t,x)$ is defined in (\ref{origin}).

With the same notations in Subsection \ref{static}, one can get that: for $j=1,2,\cdots,N$,
\begin{align}\label{a-2}
\left\{
\begin{array}{cl}
\frac{\partial{e^j(t,x)}}{\partial{t}}=&\tilde{f}(e^j(t,x);e^j(t-\tau(t),x))\\
&+\Psi(t)\sum\limits_{k=1}^N\tilde{\xi}_{jk}\Gamma^1
e^k(t,x)-\Psi(t)\sum\limits_{k=1}^N\tilde{\xi}_{jk}\Gamma^2
\Delta(e^k(t,x)),\\
&~\mathrm{if}~~t\in [t_i,s_i], i=0,1,2,\cdots\\
\frac{\partial{e^j(t,x)}}{\partial{t}}=&\tilde{f}(e^j(t,x);e^j(t-\tau(t),x))\\
&+\sum\limits_{k=1}^N\xi_{jk}\Gamma^1
e^k(t,x)-\sum\limits_{k=1}^N\xi_{jk}\Gamma^2
\Delta(e^k(t,x)),\\
&~\mathrm{if}~~t\in (s_i,t_{i+1}), i=0,1,2,\cdots
\end{array}
\right.
\end{align}

Here, the adaptive coupling strength $\Psi(t)$ is defined by (\ref{good}) with the adaptive rule
\begin{align}\label{a-3}
\dot{\Psi}(t)=\psi\max\limits_{-{\tau}\leq \varsigma\leq 0}\|e(t+\varsigma,\cdot)\|^2,
\end{align}
where $\psi>0, t\in [t_i, s_i], i=0,1,2,\cdots$, and $\|e(t,\cdot)\|$ can be found in (\ref{norm}).

Now, applying Lemma \ref{ht} to the adaptive network (\ref{a-1}), we can get the following result.
\begin{thm}\label{thm2}
Suppose Assumption \ref{add} holds, then coupled network (\ref{a-1}) with adaptive rules (\ref{good}) and (\ref{a-3}) can realize the complete synchronization asymptotically.
\end{thm}
The proof will be given in Appendix B.

Similarly, we can derive the following corollaries directly.
\begin{col}
For the coupled neural networks only with the adaptive state coupling,
\begin{align*}
\left\{
\begin{array}{cl}
\frac{\partial{w^1(t,x)}}{\partial{t}}=&f(w^1(t,x);w^1(t-\tau(t),x))+\Psi(t)\sum\limits_{k=1}^N\xi_{1k}\Gamma^1
w^k(t,x)\\
&+\Psi(t)\sigma\Gamma^1 (\pi(t,x)-w^1(t,x)),\\
\frac{\partial{w^j(t,x)}}{\partial{t}}=&f(w^j(t,x);w^j(t-\tau(t),x))+\Psi(t)\sum\limits_{k=1}^N\xi_{jk}\Gamma^1
w^k(t,x), j=2,\cdots,N\\
&~\mathrm{if}~~t\in [t_i,s_i], i=0,1,2,\cdots\\
\frac{\partial{w^j(t,x)}}{\partial{t}}=&f(w^j(t,x);w^j(t-\tau(t),x))+\sum\limits_{k=1}^N\xi_{jk}\Gamma^1
w^k(t,x), ~~j=1,2,\cdots,N,\\
&~\mathrm{if}~~t\in (s_i,t_{i+1}), i=0,1,2,\cdots
\end{array}
\right.
\end{align*}
where $\pi(t,x)$ is defined in (\ref{origin}). Suppose Assumption \ref{add} holds, then the above coupled network with adaptive rules (\ref{good}) and (\ref{a-3}) can realize the complete synchronization asymptotically.
\end{col}

\begin{col}
For the coupled neural networks only with the adaptive spatial diffusion coupling,
\begin{align*}
\left\{
\begin{array}{cl}
\frac{\partial{w^1(t,x)}}{\partial{t}}=&f(w^1(t,x);w^1(t-\tau(t),x))-\Psi(t)\sum\limits_{k=1}^N\xi_{1k}\Gamma^2
\Delta(w^k(t,x))\\
&-\Psi(t)\sigma\Gamma^2 \Delta(\pi(t,x)-w^1(t,x)),\\
\frac{\partial{w^j(t,x)}}{\partial{t}}=&f(w^j(t,x);w^j(t-\tau(t),x))-\Psi(t)\sum\limits_{k=1}^N\xi_{jk}\Gamma^2
\Delta(w^k(t,x)), \\
&~\mathrm{if}~~t\in [t_i,s_i], i=0,1,2,\cdots, j=2,3,\cdots,N;\\
\frac{\partial{w^j(t,x)}}{\partial{t}}=&f(w^j(t,x);w^j(t-\tau(t),x))-\sum\limits_{k=1}^N\xi_{jk}\Gamma^2
\Delta(w^k(t,x)), \\
&~\mathrm{if}~~t\in (s_i,t_{i+1}), i=0,1,2,\cdots, j=1,2,\cdots,N.
\end{array}
\right.
\end{align*}
where $\pi(t,x)$ is defined in (\ref{origin}). Suppose Assumption \ref{add} holds, then the above coupled network with adaptive rules (\ref{good}) and (\ref{a-3}) can realize the complete synchronization asymptotically.
\end{col}
\section{Numerical simulations}\label{nu}
To demonstrate the effectiveness of obtained theoretical results, in this section we will present some numerical simulations.

\begin{figure}
\begin{center}
\includegraphics[width=0.9\textwidth,height=0.4\textheight]{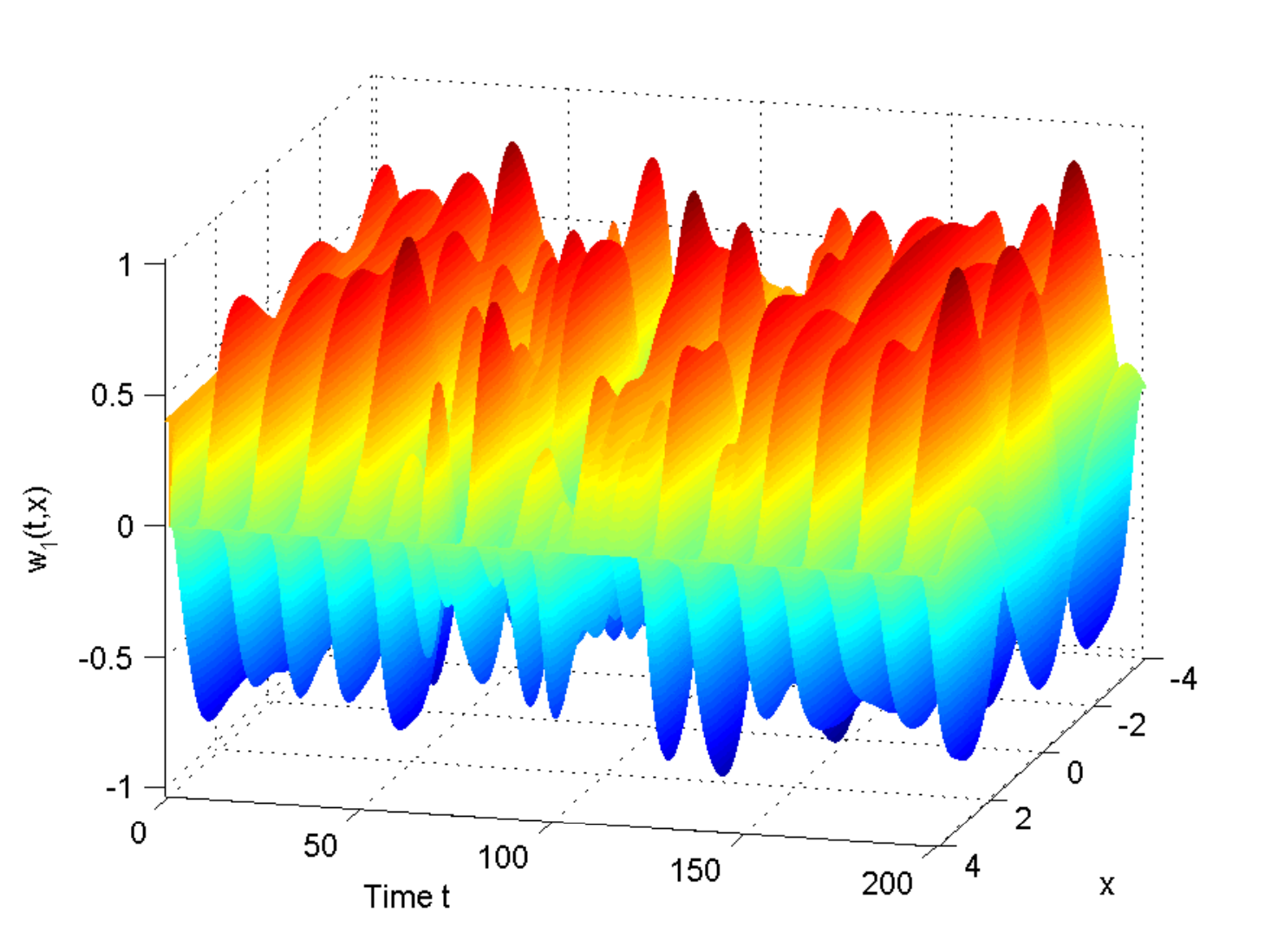}
\end{center}
\caption{Dynamical behavior of $w_1(t,x)$ for neural network (\ref{nuncoupled}) with the initial values $w(t,x)=(0.4,0.6)^T, (t,x)\in [-\tau,0]\times\Omega$}\label{wf1}
\end{figure}

\begin{figure}
\begin{center}
\includegraphics[width=0.9\textwidth,height=0.4\textheight]{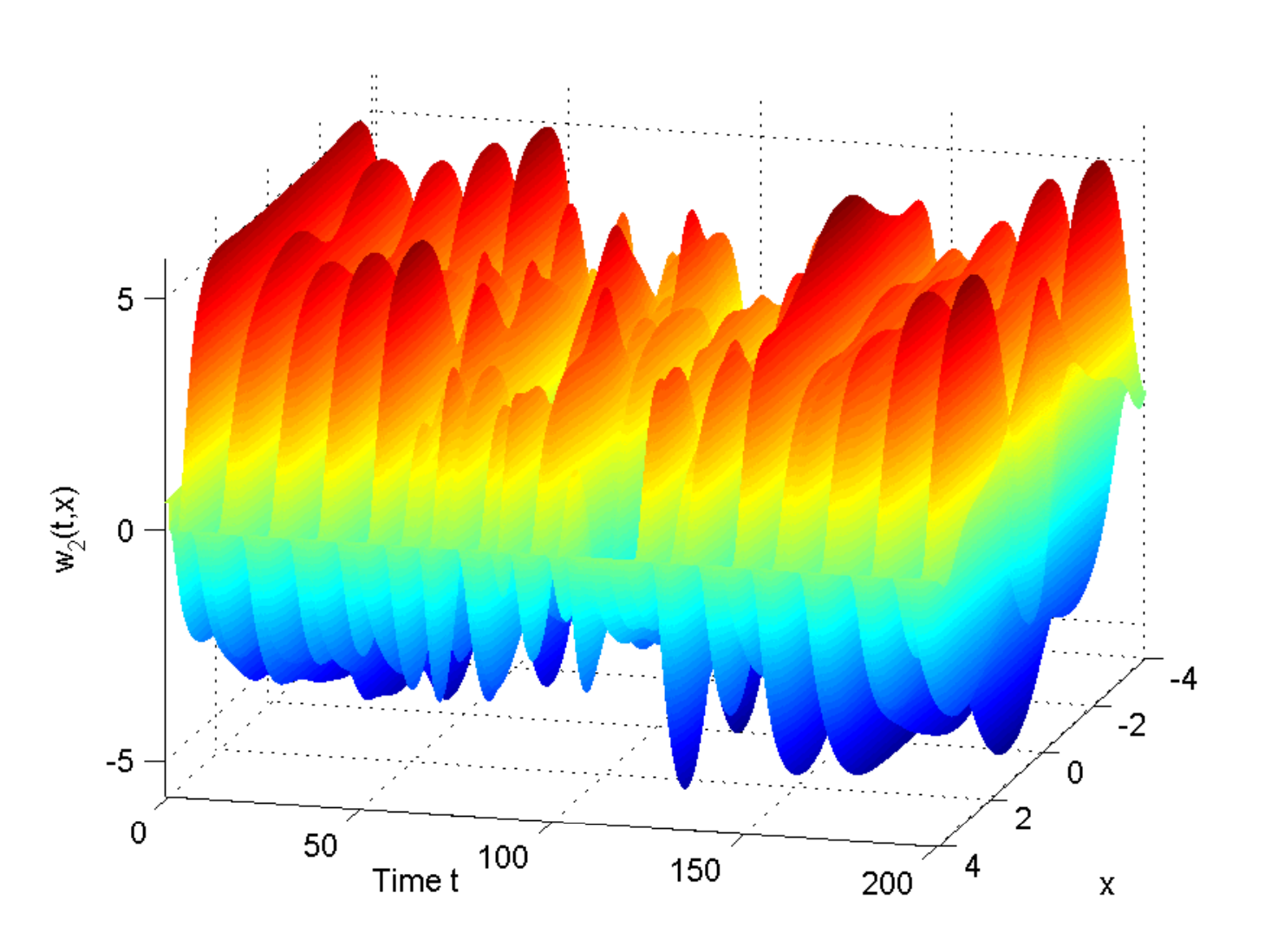}
\end{center}
\caption{Dynamical behavior of $w_2(t,x)$ for neural network (\ref{nuncoupled}) with the initial values $w(t,x)=(0.4,0.6)^T, (t,x)\in [-\tau,0]\times\Omega$}\label{wf2}
\end{figure}

\begin{figure}
\begin{center}
\includegraphics[width=0.9\textwidth,height=0.4\textheight]{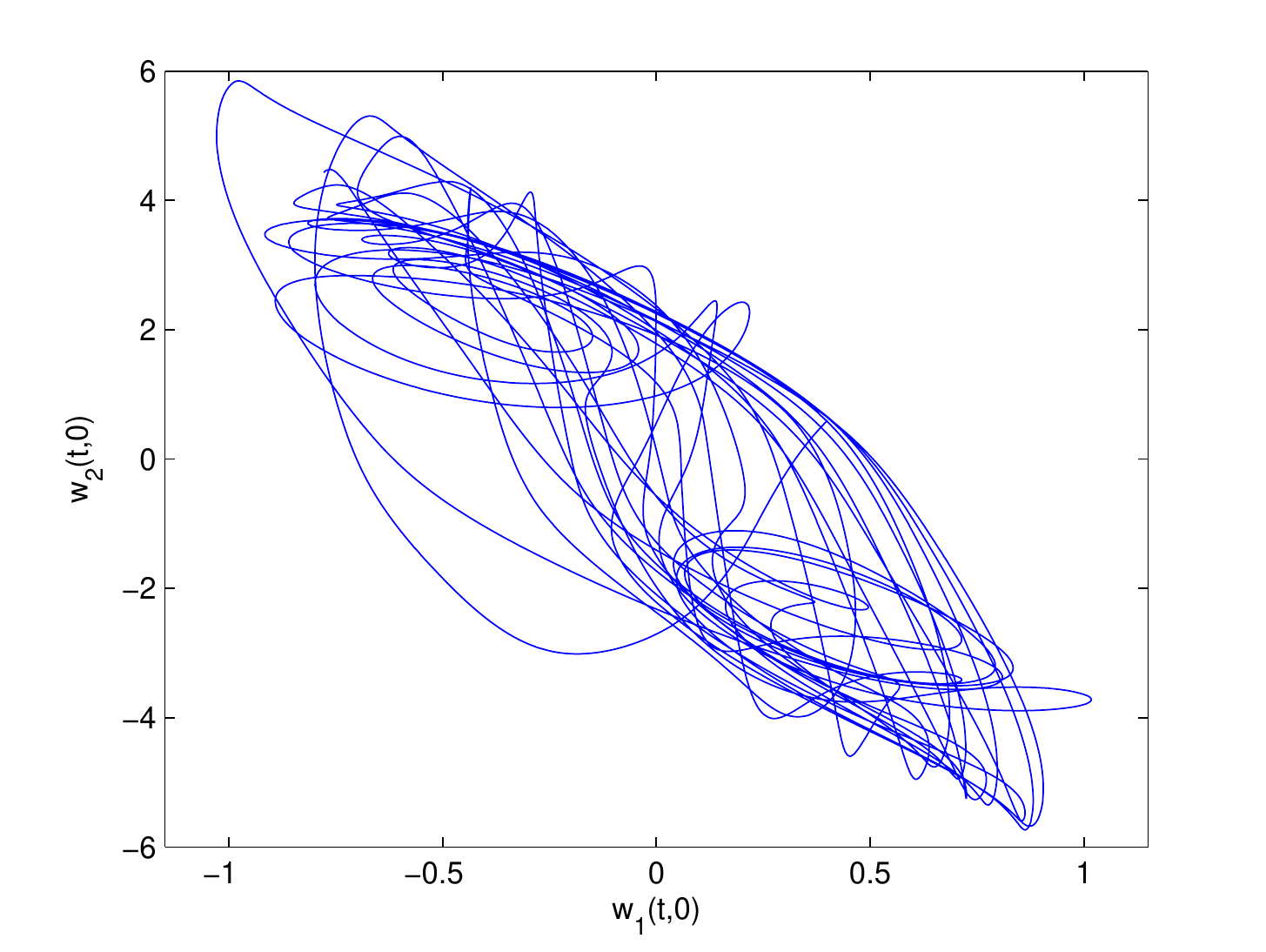}
\end{center}
\caption{The projection of neural network (\ref{nuncoupled}) on the plane $w_1(t,0)-w_2(t,0)$ with the initial values $w(t,x)=(0.4,0.6)^T, (t,x)\in [-\tau,0]\times\Omega$}\label{wf3}
\end{figure}

For the uncoupled neural network $\frac{\partial{w}(t,x)}{\partial{t}}=f(w(t,x);w(t-\tau(t),x))$, we choose the following model:
\begin{align}\label{nuncoupled}
\left\{
\begin{array}{ccl}
\frac{\partial{w}_j(t,x)}{\partial{t}}
&=&0.1\frac{\partial^2{w}_j(t,x)}{\partial{x}^2}
-c_jw_j(t,x)+\sum\limits_{k=1}^2a_{jk}g_k(w_k(t,x))\\
&&+\sum\limits_{k=1}^2b_{jk}h_k(w_k(t-\tau(t),x)), \qquad\hfill{} x\in \Omega=[-4,4] \\
w(t,x)&=&0,\quad\hfill{} (t,x)\in [-\tau,+\infty)\times\partial{\Omega}
\end{array}\right.
\end{align}
where $w(t,x)=(w_1(t,x),w_2(t,x))^T$, $j=1, 2$, and $g_k(w_k(t,x))=h_k(w_k(t,x))=\tanh(w_k(t,x)), k=1,2$, simple calculations show that Assumption \ref{add} holds with $g^{\star}=h^{\star}=1$. The time-varying delay is choose as $\tau(t)=1.1+0.2\sin t$, therefore, $\tau(t)\le \tau=1.3$; matrices $C, A, B$ are defined by
\begin{align*}
C=\left(\begin{array}{cc}1&0\\0&1\end{array}\right),~
A=\left(\begin{array}{cc}2&-0.1\\-5&3\end{array}\right),~
B=\left(\begin{array}{cc}-1.5&-0.1\\-0.2&-2.5\end{array}\right).
\end{align*}
 Figure \ref{wf1} and Figure \ref{wf2} show the dynamical behaviors of $w_1(t,x)$ and $w_2(t,x)$ for neural network (\ref{nuncoupled}) respectively; moreover, by projecting on the plane $w_1(t,0)-w_2(t,0)$, Figure \ref{wf3} shows its chaotic behavior explicitly.

\subsection{Complete synchronization in a network with a static coupling strength}\label{n-static}
Consider a linearly coupled network with four nodes as follows:
\begin{align}\label{n-1}
\left\{
\begin{array}{cl}
\frac{\partial{w^1(t,x)}}{\partial{t}}=&f(w^1(t,x);w^1(t-\tau(t),x))\\
&+c\sum\limits_{k=1}^4\xi_{1k}
w^k(t,x)-c\sum\limits_{k=1}^4\xi_{1k}
\Delta(w^k(t,x))\\
&+c\sigma (\pi(t,x)-w^1(t,x))-c\sigma \Delta(\pi(t,x)-w^1(t,x))\\
&~\mathrm{if}~~t\in [t_i,s_i], i=0,1,2,\cdots\\
\frac{\partial{w^j(t,x)}}{\partial{t}}=&f(w^j(t,x);w^j(t-\tau(t),x)),\\
&+c\sum\limits_{k=1}^4\xi_{jk}
w^k(t,x)-c\sum\limits_{k=1}^N\xi_{jk}
\Delta(w^k(t,x)),\\
&j=1,2,3,4
\end{array}
\right.
\end{align}
where $\sigma=2$, and $\pi(t,x)$ is defined by $\frac{\partial{\pi}(t,x)}{\partial{t}}=f(\pi(t,x);\pi(t-\tau(t),x))$,
where the function $f(\cdot;\cdot)$ is defined by (\ref{nuncoupled}).

As for the initial values of each node, we let $\pi(t,x)=(0.4,0.6)^T, w^1(t,x)=(0.5, 0.8)^T, w^2(t,x)=(0.6,0.5)^T, w^3(t,x)=(0.8,0.3)^T, w^4(t,x)=(0.45,0.2)^T$, where $(t,x)\in [-\tau,0]\times\Omega$.

As for the coupling matrix, we choose
\begin{align*}
\Xi=\left(\begin{array}{cccc}-2&1&0&1\\1&-2&1&0\\1&0&-2&1\\0&1&1&-2\end{array}\right),
\end{align*}
therefore, the left eigenvector $p$ corresponding to the zero eigenvalue of matrix $\Xi$ can be obtained as:
$p=(1,1,1,1)^T$.

As for the aperiodically intermittent control scheme, we choose the control time as
\begin{align*}
[0,4.9]\cup [5,9.92]\cup [9.99,14.89]\cup [14.92,19.85]\cup [19.9,24.83]\cup [24.87,29.78]\\
\cup [29.84,34.8]\cup
[34.82,39.78]\cup [39.8,44.73]\cup [44.79,49.73]\cup [49.78,54.7]\cup \cdots
\end{align*}
Obviously, we have $\omega=5$ and $\theta=4.9$ defined in Assumption \ref{asutime}, and $\rho^{\star}=0.02$ defined in (\ref{max}).

Choose $\varepsilon=6.0989, \varepsilon_2=0.5$ and the coupling strength $c=250$, then the parameters defined in (\ref{wa1})-(\ref{w8}) can be derived as: $d=0.0063$, $\alpha_1=46.5607$, $\alpha_2=1$, $\alpha_3=-150.175$, $\alpha_4=-9.375$, $\beta_1=113.0019$, $\beta_3=46.5481$, $\beta=159.55$, $\lambda=3.6115$, and $\delta=0.4205>0$. Therefore, conditions in Theorem \ref{thm1} are satisfied, i.e., the complete synchronization can be realized.

\begin{figure}
\begin{center}
\includegraphics[width=\textwidth,height=0.5\textheight]{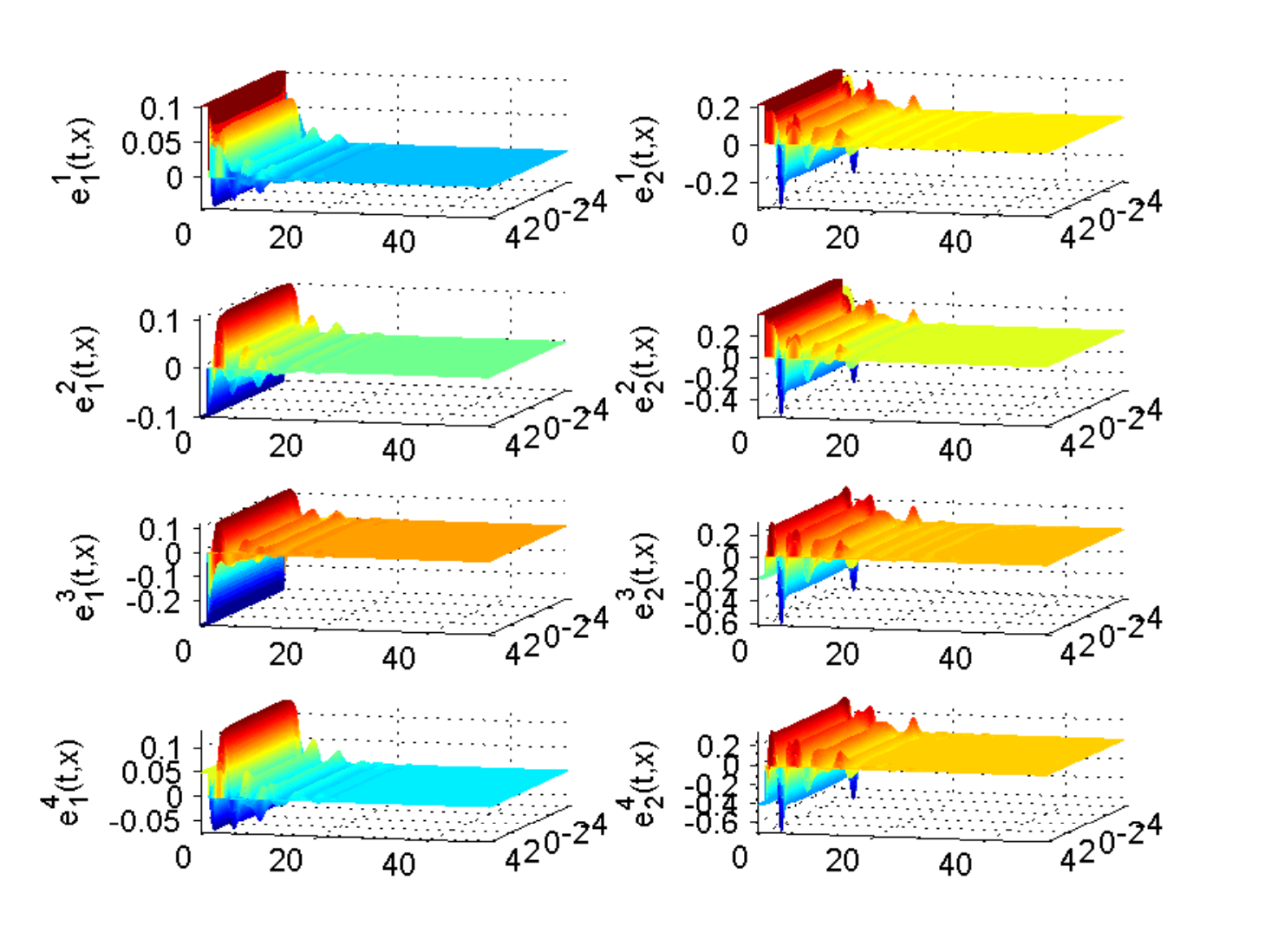}
\end{center}
\caption{Dynamical behavior of error $e^j_k(t,x), j=1,2,3,4; k=1,2$ when the coupling strength $c=3.5$}\label{wf4}
\end{figure}

\begin{figure}
\begin{center}
\includegraphics[width=0.9\textwidth,height=0.35\textheight]{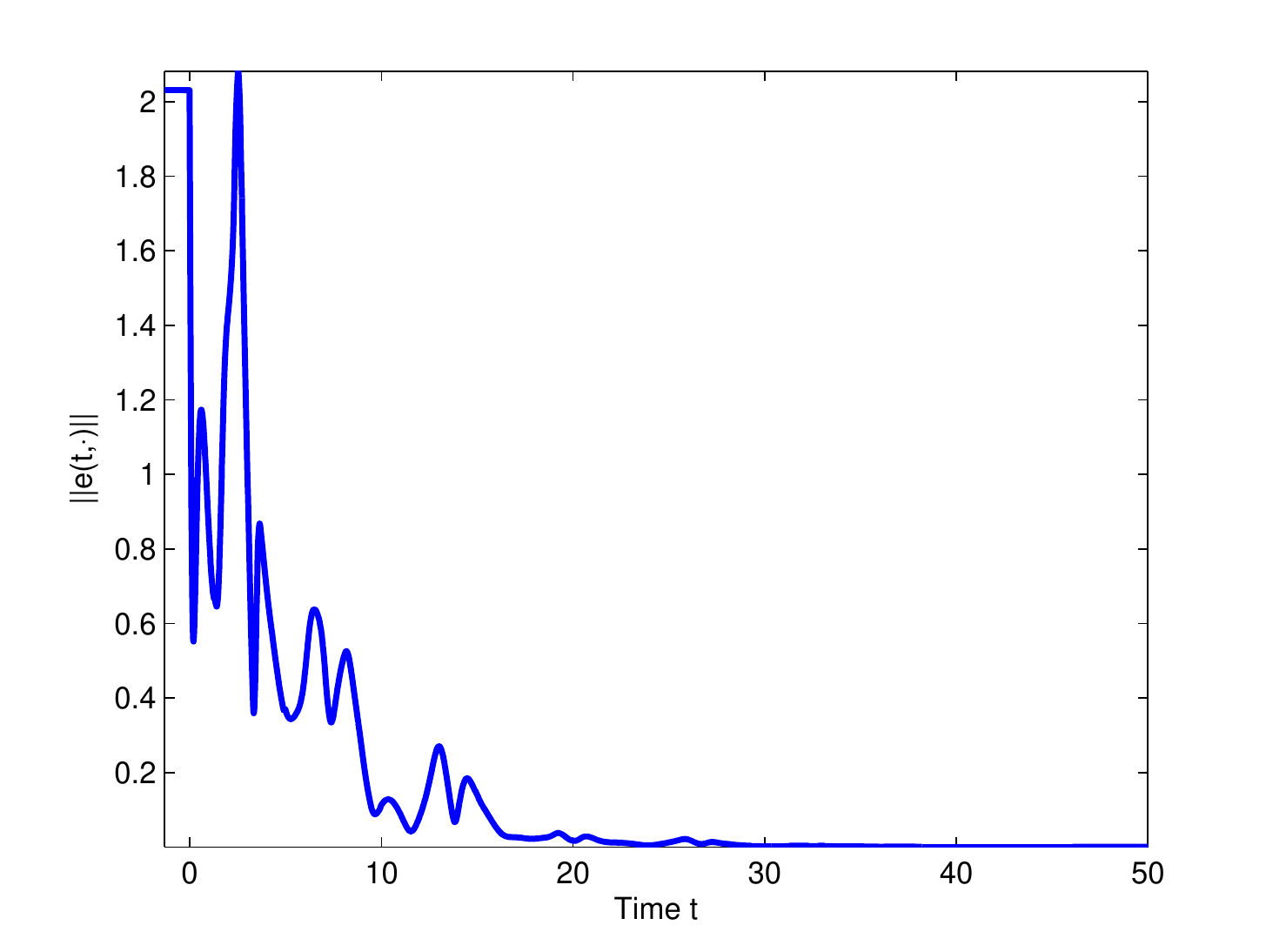}
\end{center}
\caption{Dynamical behavior of network error $\|e(t,\cdot)\|$ defined in (\ref{norm}) when the coupling strength $c=3.5$}\label{wf5}
\end{figure}

Using the Crank-Nicolson method for PDEs, we can simulate the corresponding numerical examples. In fact, the coupling strength for the complete synchronization can be smaller than the calculated value, and simulations show that if $c\ge 3$, then the complete synchronization can be realized. Figure \ref{wf4} and Figure \ref{wf5} illustrate the dynamical behaviors of error $e^j_k(t,x)=x^j_k(t,x)-\pi(t,x), j=1,2,3,4; k=1,2$ and network error $\|e(t,\cdot)\|$ defined in (\ref{norm}) when the coupling strength $c=3.5$.

\subsection{Complete synchronization with an adaptive coupling strength}
In this subsection, we consider the following network with an adaptive coupling strength:
\begin{align}\label{n-2}
\left\{
\begin{array}{cl}
\frac{\partial{w^1(t,x)}}{\partial{t}}=&f(w^1(t,x);w^1(t-\tau(t),x))\\
&+\Psi(t)\sum\limits_{k=1}^4\xi_{1k}
w^k(t,x)-\Psi(t)\sum\limits_{k=1}^4\xi_{1k}
\Delta(w^k(t,x))\\
&+2\Psi(t)(\pi(t,x)-w^1(t,x))-2\Psi(t)\Delta(\pi(t,x)-w^1(t,x)),\\
\frac{\partial{w^j(t,x)}}{\partial{t}}=&f(w^j(t,x);w^j(t-\tau(t),x))\\
&+\Psi(t)\sum\limits_{k=1}^4\xi_{jk}
w^k(t,x)-\Psi(t)\sum\limits_{k=1}^4\xi_{jk}
\Delta(w^k(t,x)), j=2,3,4,\\
&~\mathrm{if}~~t\in [t_i,s_i], i=0,1,2,\cdots\\
\frac{\partial{w^j(t,x)}}{\partial{t}}=&f(w^j(t,x);w^j(t-\tau(t),x))\\
&+\sum\limits_{k=1}^4\xi_{jk}
w^k(t,x)-\sum\limits_{k=1}^4\xi_{jk}
\Delta(w^k(t,x)), ~j=1,2,3,4,\\
&~\mathrm{if}~~t\in (s_i,t_{i+1}), i=0,1,2,\cdots
\end{array}
\right.
\end{align}
where all the parameters are the same with those in Subsection \ref{n-static}.

To illustrate the effect of aperiodically intermittent control more clearly, here we choose the control time span as
\begin{align*}
&[0,3]\cup [5,9]\cup [9.5,13]\cup [14,18]\cup [18.3,22]\cup [23,26.5]\\
\cup &[27,31]\cup
[31.7,35]\cup [36,40.5]\cup [41,45.2]\cup [45.9,50]\cup \cdots
\end{align*}
In this case, $\omega=5$ and $\theta=3$, since $3>1.3=\tau$, therefore, (\ref{smalltau}) holds.

\begin{figure}
\begin{center}
\includegraphics[width=\textwidth,height=0.45\textheight]{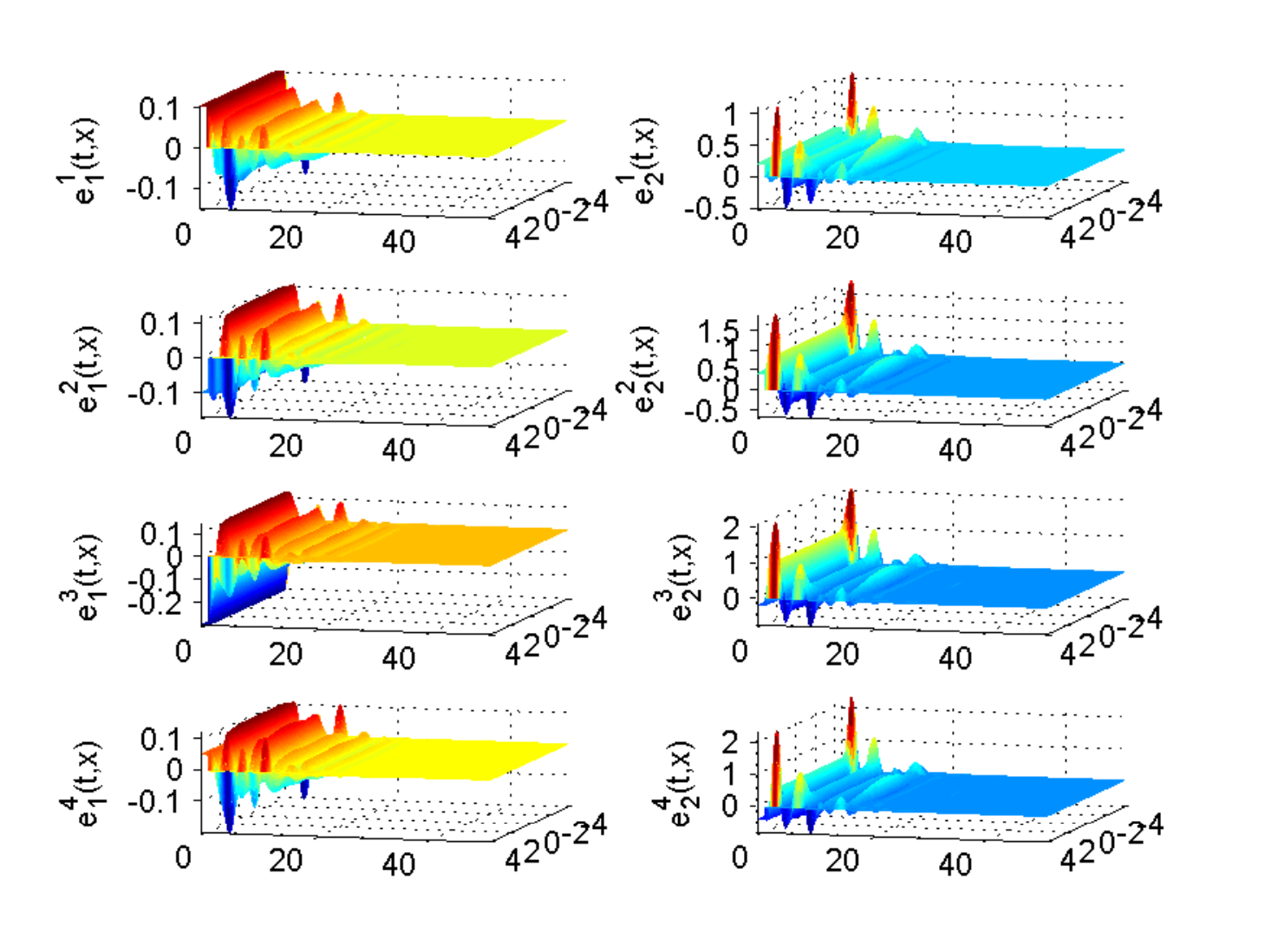}
\end{center}
\caption{Dynamical behavior of error $e^j_k(t,x), j=1,2,3,4; k=1,2$ under the adaptive rule (\ref{n-ada})}\label{wf6}
\end{figure}

\begin{figure}
\begin{center}
\includegraphics[width=0.9\textwidth,height=0.4\textheight]{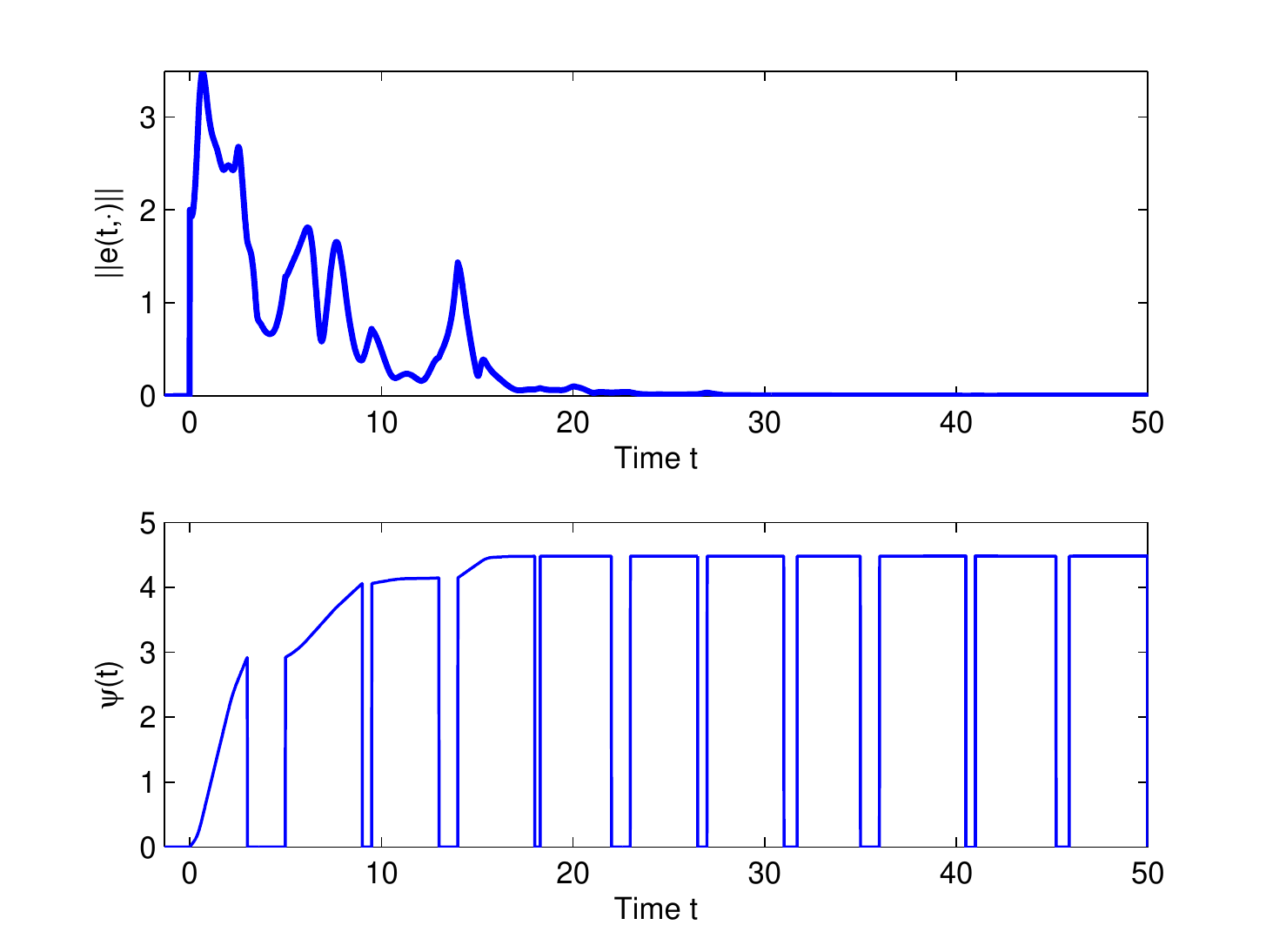}
\end{center}
\caption{Dynamical behavior of network error $\|e(t,\cdot)\|$ defined in (\ref{norm}) and the coupling strength $\Psi(t)$ under the adaptive rule (\ref{n-ada})}\label{wf7}
\end{figure}

We design the adaptive rules as
\begin{align}\label{n-ada}
\dot{\Psi}(t)=0.1\max\limits_{-{\tau}\leq \varsigma\leq 0}\|e(t+\varsigma,\cdot)\|^2, t\in [t_i,s_i], i=0,1,\cdots
\end{align}
Then, according to Theorem \ref{thm2}, the complete synchronization can be realized. Under the above adaptive scheme, Figure \ref{wf6} shows the dynamical behaviors of $e^j_k(t,x)=x^j_k(t,x)-\pi(t,x), j=1,2,3,4; k=1,2$; while Figure \ref{wf7} illustrates the dynamics of network error $\|e(t,\cdot)\|$ defined in (\ref{norm}) and the coupling strength $\Psi(t)$.

\section{Conclusion}\label{conclude}
In this paper, the complete synchronization for linearly coupled neural networks with time-varying delays and reaction-diffusion terms under aperiodically intermittent control is investigated. We first propose a novel spatial coupling protocol for the synchronization, then present some criteria for the complete synchronization with no constraint on the time-varying delay. Moreover, for the small delay case, i.e., when the bound of time-varying delay is less than the infimum of the control span, we propose a simple adaptive rule for the coupling strength, and prove its validity rigorously. Finally, some simulations are given to demonstrate the effectiveness of obtained theoretical results.

\section*{Appendix A: Proof of Theorem \ref{thm1}}\label{appA}
\begin{proof}
Define a Lyapunov function as follows:
\begin{align}\label{lyap_1}
V(t)=\frac{1}{2}\sum\limits_{j=1}^N\int_{\Omega}p^je^j(t,x)^Te^j(t,x)dx=\frac{1}{2}\int_{\Omega}e^T(t,x)(P\otimes I)e(t,x)dx,
\end{align}
where $P=\mathrm{diag}(p^1,\cdots,p^N)$ is a definite positive diagonal matrix defined in Lemma \ref{lyasta}, and $e(t,x)=(e^1(t,x)^T,\cdots,e^N(t,x)^T)^T$.

When $t\in [t_i, s_i], i=0,1,2,\cdots$, differentiating $V(t)$, we have
\begin{align}\label{v0}
&\dot{V}(t)=\sum\limits_{j=1}^N\int_{\Omega}p^je^j(t,x)^T\frac{\partial{e}^j(t,x)}{\partial{t}}dx\nonumber\\
=&\sum\limits_{j=1}^N\int_{\Omega}p^je^j(t,x)^T\bigg[f^{rd}(e^j(t,x))-Ce^j(t,x)+A\tilde{g}(e^j(t,x))+B\tilde{h}(e^j(t-\tau(t),x))\nonumber\\
&+\sum\limits_{k=1}^N\tilde{\xi}_{jk}\Gamma^1
e^k(t,x)-\sum\limits_{k=1}^N\tilde{\xi}_{jk}\Gamma^2
\Delta(e^k(t,x))\bigg]dx\nonumber\\
=&V_1+V_2+V_3+V_4+V_5+V_6.
\end{align}

From Green's formula, the Dirichlet boundary
condition, and Lemma \ref{use}, we have
\begin{align}
&\int_{\Omega}e^j(t,x)^Tf^{rd}  (e^j(t,x))dx=\sum\limits_{k=1}^n\int_{\Omega}e^j_k(t,x)\sum\limits_{r=1}^m
\frac{\partial}{\partial{x}_r}\bigg(D_{kr}\frac{\partial{e}^j_k(t,x)}{\partial{x}_r}
\bigg)dx\nonumber\\
=&\sum\limits_{k=1}^n\int_{\Omega}e^j_k(t,x)\nabla\cdot\bigg(D_{kr}
\frac{\partial{e}^j_k(t,x)}{\partial{x}_r}\bigg)
_{r=1}^mdx\nonumber\\
=&\sum\limits_{k=1}^n\int_{\Omega}\nabla\cdot\bigg(e^j_k(t,x)D_{kr}
\frac{\partial{e}^j_k(t,x)}{\partial{x}_r}\bigg)
_{r=1}^mdx\nonumber\\
&-\sum\limits_{k=1}^n\int_{\Omega}
\bigg(D_{kr}\frac{\partial{e}^j_k(t,x)}{\partial{x}_r}\bigg)
_{r=1}^m\nabla\cdot e^j_k(t,x)dx\nonumber\\
=&\sum\limits_{k=1}^n\int_{\partial{\Omega}}
\bigg(e^j_k(t,x)D_{kr}\frac{\partial{e}^j_k(t,x)}{\partial{x}_r}\bigg)
_{r=1}^mdx-\sum\limits_{k=1}^n\sum\limits_{r=1}^m\int_{\Omega}D_{kr}
\bigg(\frac{\partial{e}^j_k(t,x)}{\partial{x}_r}
\bigg)^2dx\nonumber\\
=&-\sum\limits_{k=1}^n\sum\limits_{r=1}^m\int_{\Omega}D_{kr}
\bigg(\frac{\partial{e}^j_k(t,x)}{\partial{x}_r}
\bigg)^2dx\leq -\sum\limits_{k=1}^n\sum\limits_{r=1}^m\int_{\Omega}\frac{D_{kr}}{l_r^2}
{{e}^j_k(t,x)}^2dx\nonumber\\
=&-\sum\limits_{k=1}^n\int_{\Omega}\sum\limits_{r=1}^m\frac{D_{kr}}{l_r^2}
{{e}^j_k(t,x)}^2dx\le -d\sum\limits_{k=1}^n\int_{\Omega}
{{e}^j_k(t,x)}^2dx,\nonumber
\end{align}
therefore,
\begin{align}\label{v1}
V_1=&\sum\limits_{j=1}^N\int_{\Omega}p^je^j(t,x)^Tf^{rd}(e^j(t,x))dx\nonumber\\
\le &-d\sum\limits_{j=1}^N\int_{\Omega}p^je^j(t,x)^Te^j(t,x)dx=-2dV(t).
\end{align}

According to Assumption \ref{add}, one can get
\begin{align*}
&\int_{\Omega}e^j(t,x)^T\bigg[-Ce^j(t,x)+A\tilde{g}(e^j(t,x))+B\tilde{h}(e^j(t-\tau(t),x))\bigg]dx\\
=&-C\int_{\Omega}e^j(t,x)^Te^j(t,x)dx\\
&+\int_{\Omega}e^j(t,x)^TA\tilde{g}(e^j(t,x))dx+\int_{\Omega}e^j(t,x)^TB\tilde{h}(e^j(t-\tau(t),x))dx\\
\le&\int_{\Omega}e^j(t,x)^T(-C)e^j(t,x)dx+\int_{\Omega}e^j(t,x)^T[\varepsilon_1^{-1}AA^T+\varepsilon_1{g^{\star}}^2]e^j(t,x)dx\\
&+\int_{\Omega}e^j(t,x)^T(\varepsilon_2^{-1}BB^T)e^j(t,x)dx+\varepsilon_2{h^{\star}}^2\int_{\Omega}e^j(t-\tau(t),x)^Te^j(t-\tau(t),x)dx\\
=&\int_{\Omega}e^j(t,x)^T(-C+\varepsilon_1^{-1}AA^T+\varepsilon_1{g^{\star}}^2+\varepsilon_2^{-1}BB^T)e^j(t,x)dx\\
&+\varepsilon_2{h^{\star}}^2\int_{\Omega}e^j(t-\tau(t),x)^Te^j(t-\tau(t),x)dx\\
\le&\frac{\alpha_1}{2}\int_{\Omega}e^j(t,x)^Te^j(t,x)dx+\varepsilon_2{h^{\star}}^2\int_{\Omega}e^j(t-\tau(t),x)^Te^j(t-\tau(t),x)dx,
\end{align*}
therefore,
\begin{align}\label{v234}
&V_2+V_3+V_4\nonumber\\
=&\sum\limits_{j=1}^Np^j\int_{\Omega}e^j(t,x)^T\bigg[-Ce^j(t,x)+A\tilde{g}(e^j(t,x))+B\tilde{h}(e^j(t-\tau(t),x))\bigg]dx\nonumber\\
\le&\alpha_1V(t)+\alpha_2V(t-\tau(t)).
\end{align}

According to Lemma \ref{lyasta}, $\{P\tilde{\Xi}\}^s$ is negative definite, we have
\begin{align}\label{v5}
V_5=&\sum\limits_{j=1}^N\int_{\Omega}p^je^j(t,x)^T\sum\limits_{k=1}^N\tilde{\xi}_{jk}\Gamma^1
e^k(t,x)=\int_{\Omega}e(t,x)^T(\{P\tilde{\Xi}\}^s\otimes \Gamma^1)e(t,x)dx\nonumber\\
\le&\min_{k=1,\cdots,n}\gamma_k^1\int_{\Omega}e(t,x)^T(\{P\tilde{\Xi}\}^s\otimes I)e(t,x)dx\nonumber\\
\le&\min_{k=1,\cdots,n}\gamma_k^1\lambda_{\max}(\{P\tilde{\Xi}\}^s)\int_{\Omega}e(t,x)^Te(t,x)dx\nonumber\\
\le&2\min_{k=1,\cdots,n}\gamma_k^1\lambda_{\max}(\{P\tilde{\Xi}\}^s)V(t)=\alpha_3V(t).
\end{align}

With the same reason, one can also get that
\begin{align}\label{v6}
V_6=&-\sum\limits_{j=1}^N\int_{\Omega}p^je^j(t,x)^T\sum\limits_{k=1}^N\tilde{\xi}_{jk}\Gamma^2
\Delta(e^k(t,x))dx\nonumber\\
=&-\sum\limits_{j=1}^N\sum\limits_{k=1}^Np^j\tilde{\xi}_{jk}\int_{\Omega}e^j(t,x)^T\Gamma^2
\Delta(e^k(t,x))dx\nonumber\\
=&\sum_{r=1}^m\sum\limits_{j=1}^N\sum\limits_{k=1}^Np^j\tilde{\xi}_{jk}\sum_{q=1}^n\gamma^2_{q}\int_{\Omega}\frac{\partial{e^j_q(t,x)}}{\partial{x_r}}
\frac{\partial{e^k_q(t,x)}}{\partial{x_r}}dx\nonumber\\
=&\sum_{r=1}^m\int_{\Omega}\bigg(\frac{\partial{e(t,x)}}{\partial{x_r}}\bigg)^T
(\{P\tilde{\Xi}\}^s\otimes \Gamma^2)\frac{\partial{e(t,x)}}{\partial{x_r}}dx\nonumber\\
\le&\min_{k=1,\cdots,n}\gamma_k^2\lambda_{\max}(\{P\tilde{\Xi}\}^s)\sum_{r=1}^m\int_{\Omega}\bigg(\frac{\partial{e(t,x)}}{\partial{x_r}}\bigg)^T
\frac{\partial{e(t,x)}}{\partial{x_r}}dx\nonumber\\
\le&\min_{k=1,\cdots,n}\gamma_k^2\lambda_{\max}(\{P\tilde{\Xi}\}^s)\sum_{r=1}^m\frac{1}{l_r^2}\int_{\Omega}e(t,x)^Te(t,x)dx\nonumber\\
\le&2\min_{k=1,\cdots,n}\gamma_k^2\lambda_{\max}(\{P\tilde{\Xi}\}^s)\sum_{r=1}^m\frac{1}{l_r^2}V(t)=\alpha_4V(t).
\end{align}

Therefore, combining with (\ref{v0})-(\ref{v6}), one can get that
\begin{align}\label{control}
\dot{V}(t)\leq -\beta_1V(t)+\alpha_2V(t-\tau(t)).
\end{align}

On the other hand, for $t\in (s_i,t_{i+1}), i=0,1,2,\cdots$, recalling the fact that $\{P\Xi\}^s$ is a symmetric matrix with its column and row sums being zero, so $\lambda_{\max}(\{P\Xi\}^s)=0$, by the similar analysis, we have
\begin{align}\label{rest}
\dot{V}(t)\leq \beta_3V(t)+\alpha_2V(t-\tau(t)).
\end{align}

Combining with (\ref{control})-(\ref{rest}), and according to Lemma \ref{synlem}, we can deduce that the complete synchronization can be finally realized, i.e.,
$V(t)\le \sup_{-\tau\le\kappa\le}V(\kappa)e^{-\varepsilon t}$, therefore, $\|e(t,\cdot)\|=O(e^{-\varepsilon t/2})$. The proof is completed.
\end{proof}

\section*{Appendix B: Proof of Theorem \ref{thm2}}\label{appB}
\begin{proof}
From (\ref{a-2}), we can have the compact form for the error equation with $e(t,x)=(e^1(t,x)^T,\cdots,e^N(t,x)^T)^T$,
\begin{align}\label{b-1}
\left\{
\begin{array}{cl}
\frac{\partial{e(t,x)}}{\partial{t}}=&\tilde{F}(e(t,x);e(t-\tau(t),x))\\
&+\Psi(t)(\tilde{\Xi}\otimes\Gamma^1)
e(t,x)-\Psi(t)(\tilde{\Xi}\otimes\Gamma^2)
\Delta(e(t,x)),\\
&~\mathrm{if}~~t\in [t_i,s_i], i=0,1,2,\cdots\\
\frac{\partial{e(t,x)}}{\partial{t}}=&\tilde{F}(e(t,x);e(t-\tau(t),x))\\
&+({\Xi}\otimes\Gamma^1)
e(t,x)-({\Xi}\otimes\Gamma^2)
\Delta(e(t,x))\\
&~\mathrm{if}~~t\in (s_i,t_{i+1}), i=0,1,2,\cdots
\end{array}
\right.
\end{align}
where $\Delta(e(t,x))=(\Delta(e^1(t,x))^T,\cdots,\Delta(e^N(t,x))^T)^T$, and
\begin{align*}
&\tilde{F}(e(t,x);e(t-\tau(t),x))\\
=&(\tilde{f}^T(e^1(t,x);e^1(t-\tau(t),x)),\cdots,\tilde{f}^T(e^N(t,x);e^N(t-\tau(t),x))).
\end{align*}

We define a new function
\begin{align*}
&\bar{F}(e(t,x);e(t-\tau(t),x))=\\
&\left\{
\begin{array}{ll}
\tilde{F}(e(t,x);e(t-\tau(t),x)), ~~~~~~~~~~~~~~~~~~~~~~~~~~~~~~~~~~~~~~~~~~~~~\mathrm{if}~t\in[t_i,s_i],\\
\tilde{F}(e(t,x);e(t-\tau(t),x))+({\Xi}\otimes\Gamma^1)
e(t,x)-({\Xi}\otimes\Gamma^2)
\Delta(e(t,x)),\\
~~~~~~~~~~~~~~~~~~~~~~~~~~~~~~~~~~~~~~~~~~~~~~~~~~~~~~~~~~~~~~~~~~~~~~~~~~~~~~~~~~~~~~~~\mathrm{if}~t\in(s_i,t_{i+1}).
\end{array}
\right.
\end{align*}
Therefore, according to (\ref{rest}), we have
\begin{align*}
&\int_{\Omega}e(t,x)^T(P\otimes I)\bar{F}(e(t,x);e(t-\tau(t),x))dx\\
\leq&\int_{\Omega}e(t,x)^T(P\otimes I)\tilde{F}(e(t,x);e(t-\tau(t),x))dx\\
\leq&\frac{1}{2}\beta_3\int_{\Omega}e(t,x)^T(P\otimes I)e(t,x)dx+\frac{\alpha_2}{2}\int_{\Omega}e(t-\tau(t),x)^T(P\otimes I)e(t-\tau(t),x)dx
\end{align*}
where $\beta_3$ and $\alpha_2$ are defined in (\ref{w7}) and (\ref{w3}), respectively.

Define the Lyapunov function as (\ref{lyap_1}), then with the same process as the proof of Theorem \ref{thm1}, we have
\begin{align*}
\dot{V}(t)=&\int_{\Omega}e(t,x)^T(P\otimes I)\bar{F}(e(t,x);e(t-\tau(t),x))dx\\
&+\Psi(t)\int_{\Omega}e(t,x)^T(P\otimes I)\bigg((\tilde{\Xi}\otimes\Gamma^1)
e(t,x)-\Psi(t)(\tilde{\Xi}\otimes\Gamma^2)
\Delta(e(t,x))\bigg)dx\\
\leq &\beta_3V(t)+\alpha_2V(t-\tau(t))+\Psi(t)(\alpha_3+\alpha_4)V(t),
\end{align*}
where $\alpha_3<0$ and $\alpha_4<0$ are defined in (\ref{w4}) and (\ref{w5}), respectively.

According to Lemma \ref{ht}, we can obtain $\lim\limits_{t\rightarrow\infty}V(t)=0$, which implies that the complete synchronization can be realized.

The proof is completed.
\end{proof}
\section*{Acknowledgement}
This work was supported by the National Science Foundation of China under Grant Nos. 61203149, 61233016, 11471190 and 11301389;
the National Basic Research Program of China (973 Program) under Grant No. 2010CB328101; ``Chen Guang'' project supported by Shanghai Municipal Education Commission and Shanghai Education Development Foundation under Grant No. 11CG22; the Fundamental Research Funds for the Central Universities of Tongji University; the NSF of Shandong Province under Grant No. ZR2014AM002, project funded by China Postdoctoral Science Foundation under Grant Nos. 2012M511488 and 2013T60661, the Special Funds for Postdoctoral Innovative Projects of Shandong Province under Grant No. 201202023, IIF of Shandong University under Grant No. 2012TS019.

\end{document}